\documentclass[aps,showpacs,prl,reprint,longbibliography,superscriptaddress]{revtex4-1}

\usepackage[colorlinks=true,linkcolor=blue,citecolor=blue,urlcolor=blue]{hyperref}
\usepackage{setspace} %to set 1.5 spacing in text
\usepackage{graphicx}
\usepackage{amsmath}
\usepackage{color}
\usepackage{amsmath}
\usepackage{amssymb}
\usepackage{verbatim}
\usepackage{latexsym}
\usepackage{enumerate} % to change the numbering of 'enumerate'
\usepackage{bm} % for bold face mathematical symbols
\usepackage[caption=false]{subfig}
\usepackage{multirow}
\usepackage{booktabs}
\usepackage{siunitx}

\usepackage{float}

\usepackage{lipsum}

\begin{document}

\title{\textbf{\Large{The impact of exciton delocalization on exciton-vibration interactions in organic semiconductors}}}

\author{Antonios M. Alvertis}
\email{ama80@cam.ac.uk}
\affiliation{Cavendish Laboratory, University of Cambridge, J.\,J.\,Thomson Avenue, Cambridge CB3 0HE, United Kingdom}
\author{Raj Pandya}
\affiliation{Cavendish Laboratory, University of Cambridge, J.\,J.\,Thomson Avenue, Cambridge CB3 0HE, United Kingdom}
\author{Loreta A. Muscarella}
\affiliation{AMOLF, Center for Nanophotonics, Science Park 104, 1098 XG Amsterdam, The Netherlands}
\author{Nipun Sawhney}
\affiliation{Cavendish Laboratory, University of Cambridge, J.\,J.\,Thomson Avenue, Cambridge CB3 0HE, United Kingdom}
\author{Malgorzata Nguyen}
\affiliation{Cavendish Laboratory, University of Cambridge, J.\,J.\,Thomson Avenue, Cambridge CB3 0HE, United Kingdom}
\author{Bruno Ehrler}
\affiliation{AMOLF, Center for Nanophotonics, Science Park 104, 1098 XG Amsterdam, The Netherlands}
\author{Akshay Rao}
\affiliation{Cavendish Laboratory, University of Cambridge, J.\,J.\,Thomson Avenue, Cambridge CB3 0HE, United Kingdom}
\author{Richard H. Friend}
\affiliation{Cavendish Laboratory, University of Cambridge, J.\,J.\,Thomson Avenue, Cambridge CB3 0HE, United Kingdom}
\author{Alex W. Chin}
\affiliation{CNRS \& Institut des NanoSciences de Paris, Sorbonne Universit\'e, 75252 Paris Cedex 05, France}
\author{Bartomeu Monserrat}
\email{bm418@cam.ac.uk}
\affiliation{Cavendish Laboratory, University of Cambridge, J.\,J.\,Thomson Avenue, Cambridge CB3 0HE, United Kingdom}
\affiliation{Department of Materials Science and Metallurgy, University of Cambridge, 27 Charles Babbage Road, Cambridge CB3 0FS, U.K.}

\date{\today}

\begin{abstract}

Organic semiconductors exhibit properties of individual molecules and extended crystals 
simultaneously. The strongly bound excitons they host are typically described in the 
molecular limit, but excitons can delocalize over many molecules, raising the question 
of how important the extended crystalline nature is. 
Using accurate Green's function 
based methods for the electronic structure and non-perturbative finite difference methods
for exciton-vibration coupling,
we describe exciton interactions with molecular and crystal degrees of freedom concurrently. We find that the degree of exciton delocalization controls these interactions, with thermally activated crystal phonons predominantly coupling to delocalized states, and molecular quantum fluctuations predominantly coupling to localized states. Based on this picture, we quantitatively predict and interpret the temperature and pressure dependence of excitonic peaks in the acene series of organic semiconductors, which we confirm experimentally, and we develop a simple experimental protocol for probing exciton delocalization. Overall, we
provide a unified picture of exciton delocalization and vibrational effects in organic semiconductors, reconciling the complementary views of finite molecular clusters and periodic molecular solids.

%We simultaneously capture, for the first time using a first-principles approach, the respective exothermicity and endothermicity of singlet fission in pentacene and tetracene. 

\end{abstract}

\maketitle

Optoelectronic devices based on organic semiconductors (OSCs), such as light-emitting diodes (LEDs) \cite{Reineke2009} and solar cells \cite{Meng2018}, are promising candidates for technological applications. In contrast to their inorganic counterparts, OSCs host strongly bound excitons \cite{Knupfer2003,Kohler2009} that, depending on the relative spin of the electron and hole pair, form spin-zero (singlet) or spin-one (triplet) configurations. The interconversion between singlet and triplet states is of high relevance to the application of OSCs, with examples including thermally activated delayed fluorescence used in organic LEDs \cite{Uoyama2012,Reineke2014}, and singlet exciton fission which could lead to solar cells with efficiencies surpassing the Shockley-Queisser limit \cite{Smith2010,Hanna2006,Rao2017}.

The weak van der Waals interaction between molecules in OSCs has led to 
the common approximation of using isolated molecular dimers or oligomers 
to simulate the entire crystal \cite{Beljonne2013,Berkelbach2013,Yost2014,Coto2015}. Wavefunction-based 
methods are the natural way of studying such finite-sized clusters of 
molecules, providing a good description of the underlying physics in 
those systems where exciton states are localized. While triplet excitons 
are localized in most cases, the exchange interaction drives singlet 
excitons in larger molecules to delocalize over multiple monomers  
\cite{Cudazzo2015,Refaely-Abramson2017}. Therefore, OSCs can 
simultaneously exhibit features of the molecular and extended crystal 
limits depending on the exciton spin and the molecular size, calling for 
a unified picture.

Due to the mechanically soft character of organic materials, the coupling of excitons to molecular and crystal vibrations can be extremely strong and dominate the physics that makes OSCs interesting from an application point of view. For example, exciton-vibration coupling has been shown to play a central role in processes such as singlet fission and exciton transport \cite{Arago2015,Musser2015,Bakulin2016}, and several theoretical studies have successfully captured this interaction in molecular clusters
\cite{Miyata2017, Schnedermann2019, Alvertis2019}. However, these molecule-based approaches cannot capture the long-range intermolecular vibrations that lead to strong deviations of the orbital overlap between neighboring monomers 
\cite{Troisi2005,Troisi2006}. A full treatment of exciton delocalization becomes necessary to account for the strong dynamic disorder of OSCs in solid state systems, but this remains an open challenge. 

\begin{figure}[tb]
\centering
\includegraphics[width=\linewidth]{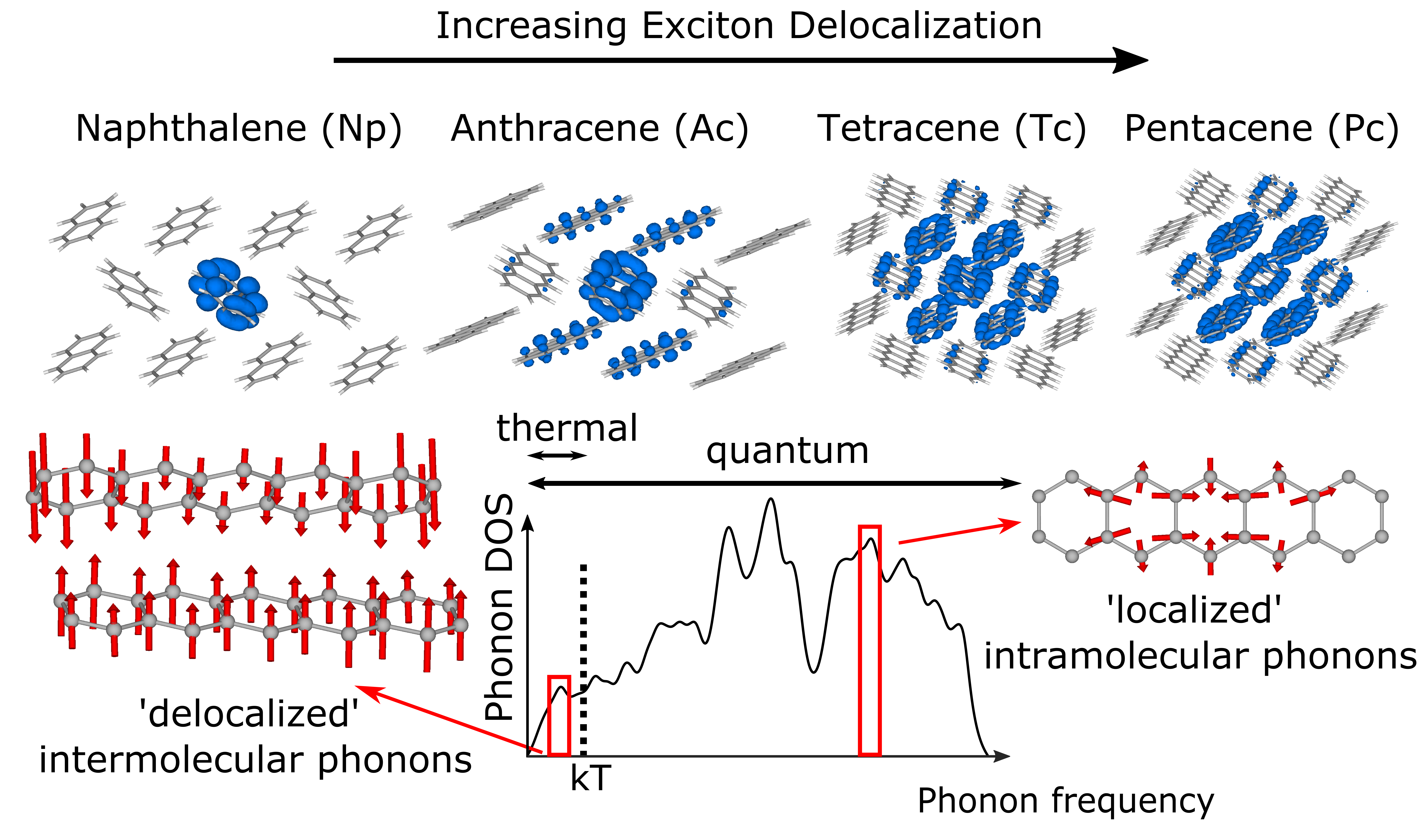}
\caption{Singlet excitons in the acene series of molecular crystals range from entirely localized (naphthalene) to highly delocalized (pentacene). The phonon density of states (DOS) of these materials consists of a thermally-activated regime of intermolecular motions, and a high-energy regime of intramolecular modes that are only active due to quantum fluctuations. The interaction of delocalized excitons is dominated by low-energy thermally-active modes, while localized excitons couple preferentially to high-energy modes via quantum fluctuations.}
\label{fig:concept}
\end{figure}

In this work, we present a theoretical framework to understand exciton-phonon 
interactions in OSCs treating molecular and crystal features on the same footing. We 
combine many-body $GW$ and Bethe-Salpeter calculations for exciton properties 
\cite{Lars1965,Rohlfing1998,Rohlfing2000}, with finite displacement methods for 
vibrational properties \cite{Kunc1982,Monserrat2018}, to capture the strong 
non-perturbative exciton-vibration interactions present in OSCs \cite{Monserrat2016_2}. 
The applied computational methodology is described in detail in the Supplemental 
Material (SM). This combination of highly accurate methods
allows us to identify the extent of exciton localization as the key parameter determining the relative importance of molecular and crystal degrees of freedom. 
To probe our theoretical framework, we use the acene series of molecular crystals because they exhibit excitons that range from entirely localized (molecular limit) to strongly delocalized (crystal limit), see Figure~\ref{fig:concept}. We find that in the molecular limit, excitons couple preferentially to high-energy phonons that are not thermally activated, so the coupling is purely via quantum fluctuations. By contrast, in the crystal limit thermally-activated low-energy phonons dominate the interaction with excitons. In intermediate cases, both contributions can be important. 

Therefore, we provide a unified picture between the molecular and crystal limits of OSCs, which has multiple implications for their properties, some examples of which we explore. Firstly, we predict that excitons in acenes should exhibit a very weak temperature dependence due to opposing effects between thermal expansion and exciton-phonon coupling driven by long-range crystal effects, which is quantitatively confirmed by our temperature dependent measurements. 
%We argue that this weak exciton peak temperature dependence should hold in the vast majority of OSCs. 
Interestingly, short-range molecular vibrations, which are negligible for the temperature dependence, lead to strong exciton energy renormalizations due to quantum fluctuations, an effect which is inherent to organic materials due to the low mass of carbon and hydrogen. Accounting for this effect provides unprecedented predictive power for \textit{absolute} exciton energies compared to experiment, something that holds particular practical value for determining the energy of triplet states, which is challenging to do experimentally \cite{Reineke2014,Ehrler2012}. 
%As a result, our calculations simultaneously describe the respective exothermicity and endothermicity of singlet fission in pentacene and tetracene for the first time purely from first principles. 
Moreover, we predict that delocalized excitons should 
exhibit stronger pressure dependencies than localized ones, and again quantitatively 
confirm this prediction with experiments in the acene series. 
%Consequently, we propose 
%that combining pressure and temperature dependent measurements is a simple experimental strategy to probe the degree of exciton delocalization in OSCs. 

%\section*{Results}

%\subsection*{Exciton delocalization}

For singlet excitonic states, molecular size determines the degree of delocalization in the solid state \cite{Cudazzo2015}. For large organic structures, the average electron-hole distance on a single molecule is comparable to the distance between neighboring molecules. Therefore, the attractive Coulomb energy between electron and hole is similar when they localize on the same or on adjacent molecules, allowing for delocalized states with large average electron-hole separation. Using the language that is commonly employed in the literature \cite{Cudazzo2015}, singlet states in these materials have a strong charge transfer (CT) character. In contrast, in small molecules the attractive Coulomb interaction of electron and hole within a single molecule is significantly stronger than if they localized on different molecules. For such small structures, this leads to an almost perfectly localized Frenkel-like singlet exciton. Triplet excitons are generally more localized than singlets, due to the lack of a repulsive exchange interaction in this state \cite{Rohlfing2000}. Therefore, depending on spin and on the size of molecules, different degrees of exciton delocalization appear in the solid state.

To demonstrate exciton delocalization, we use the acene series of OSCs, which encompasses the entire spectrum of exciton sizes. The singlet exciton wavefunctions of these crystals are visualized in Figure\,\ref{fig:concept}. The electron density is visualized in blue for a hole localized on the central monomer of the displayed area. Moving from acenes with a small number of carbon rings towards larger members of the family, the singlet becomes increasingly delocalized over several molecules, transitioning from a Frenkel-like exciton to an exciton with a strong CT character. 
%A recent systematic analysis of the degree of CT character \cite{Wang2018} of anthracene, tetracene and pentacene \cite{Liu2020} confirms our observation of increasing exciton delocalization in the acene series. 
The triplet wavefunctions of the acene crystals are visualized in SM Figure\,S1, and are highly localized as expected. 
%Throughout the rest of this paper, the term delocalization will refer to the spatial extent of the electron wavefunction for a hole localized at a specific point. 
%Since we are working with periodic crystals, the center of mass of the electrons, holes, and excitons is generally delocalized, however this is not the kind of delocalization we will be referring to from here on. 

The spectrum of exciton delocalization in crystalline OSCs suggests that excitons may respond differently to the different kinds of vibrations present in these materials. A schematic phonon density of states (DOS) for OSCs is presented in Figure\,\ref{fig:concept}, consisting of two distinct regions: a small thermally-activated regime of low-frequency phonons, corresponding to intermolecular motion, and a large region which is dominated by high-energy intramolecular motions, such as C-C stretches. While most of the phonon modes falling into the latter regime are not thermally-activated, they still oscillate with a finite energy of $\frac{1}{2}\hbar \omega$ due to zero-point quantum fluctuations, and their effect needs to be accounted for. Intuition suggests that thermally-activated modes that modulate the orbital overlap between neighboring monomers \cite{Troisi2005,Troisi2006} will predominantly have an effect on more delocalized excitons, the wavefunction of which extends over a larger number of molecules. Similarly, one expects that the high-energy intramolecular motions that are only active due to quantum fluctuations, will mostly affect localized excitons, the wavefunction of which has a greater amplitude in the vicinity of these localized motions. We now proceed to examine this picture.

%\subsection*{Thermally- and quantum-activated exciton-phonon interactions}

\begin{figure}[tb]
\centering
\includegraphics[width=\linewidth]{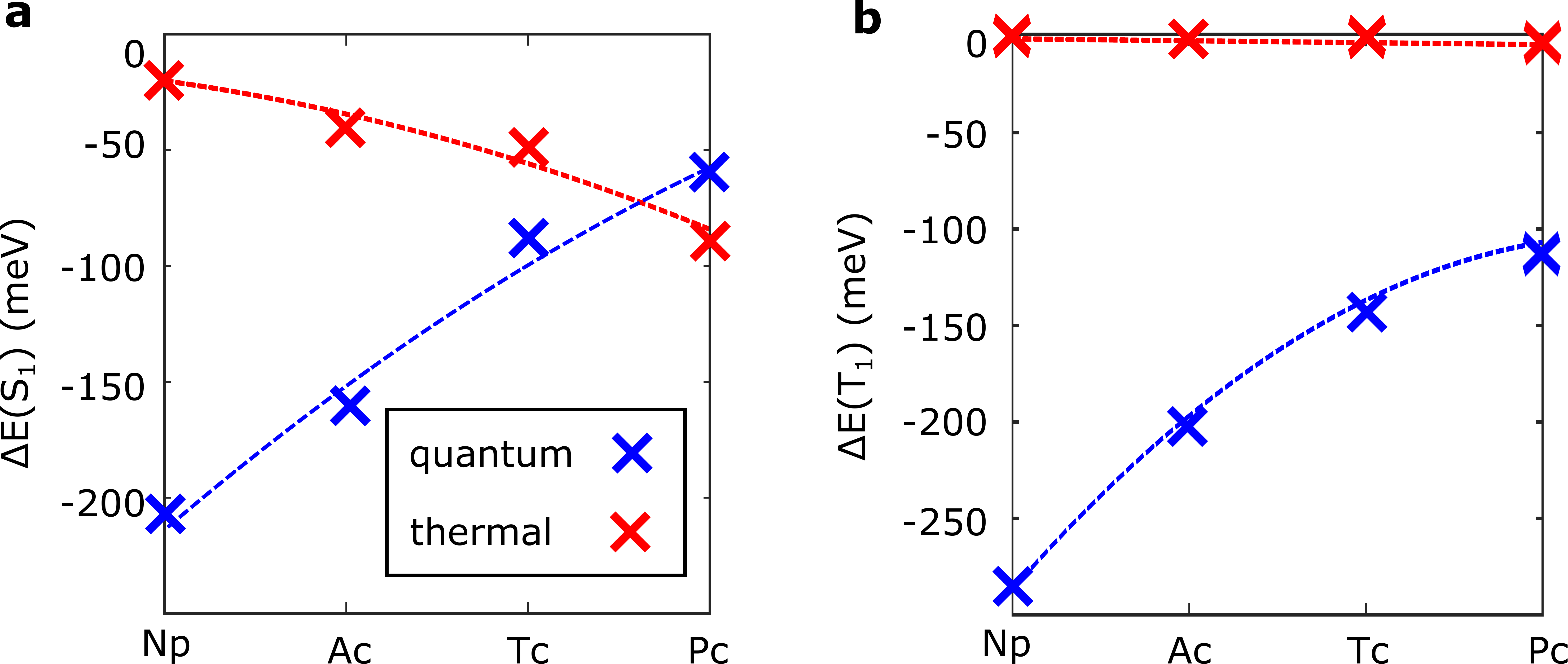}
\caption{Change caused by thermal activation (red) and quantum fluctuations (blue) of phonons to the singlet (panel \textbf{a}) and triplet (panel \textbf{b}) exciton energies of the different acenes. }
%The contribution of thermally-activated phonons at room temperature (red) is more pronounced for larger acenes with delocalized excitons, while the role of phonon quantum fluctuations (blue) is more significant for the smaller acenes with localized excitons.}
\label{fig:thermal_and_quantum}
\end{figure}

By employing the first principles computational methodology outlined in the SM, we obtain the expectation value for the singlet and triplet exciton energies of the acene crystals at $0$\,K and $300$\,K arising from exciton-phonon coupling. The interaction of the exciton with phonons leads to a red shift of its energy. This is due to the fact that the optical gap of OSCs is very large compared to typical phonon frequencies, implying that it is almost exclusively stabilizing intraband phonon-induced transitions that contribute \cite{Saha2014}.

Even at $0$\,K, phonons fluctuate with their zero-point energy, and the crystal is not ``frozen''. The interaction of the exciton with quantum fluctuations leads to a red shift of its energy compared to the static picture, which becomes more pronounced for the smaller acenes (fewer number of carbon rings), where excitons are more localized. This is visualized in Figure\,\ref{fig:thermal_and_quantum}, and comparing the singlet (panel \textbf{a}) to the triplet (panel \textbf{b}), the red shift is more important for the more localized triplet states. 

By increasing the temperature from $0$\,K to $300$\,K, low-energy intermolecular motions are activated, leading to a further red shift of exciton energies. In agreement with the intuitive picture of Figure\,\ref{fig:concept}, this effect is more important for more delocalized states, the properties of which depend on intermolecular interactions. When it comes to the perfectly localized triplet state, thermally-activated phonons have a negligible effect on its energy. 
We now proceed to use these physical arguments to uncover the mechanism behind the temperature-dependent properties of excitons. 

%\subsection*{Exciton temperature dependence}

\begin{figure}[tb]
\centering
\includegraphics[width=0.8\linewidth]{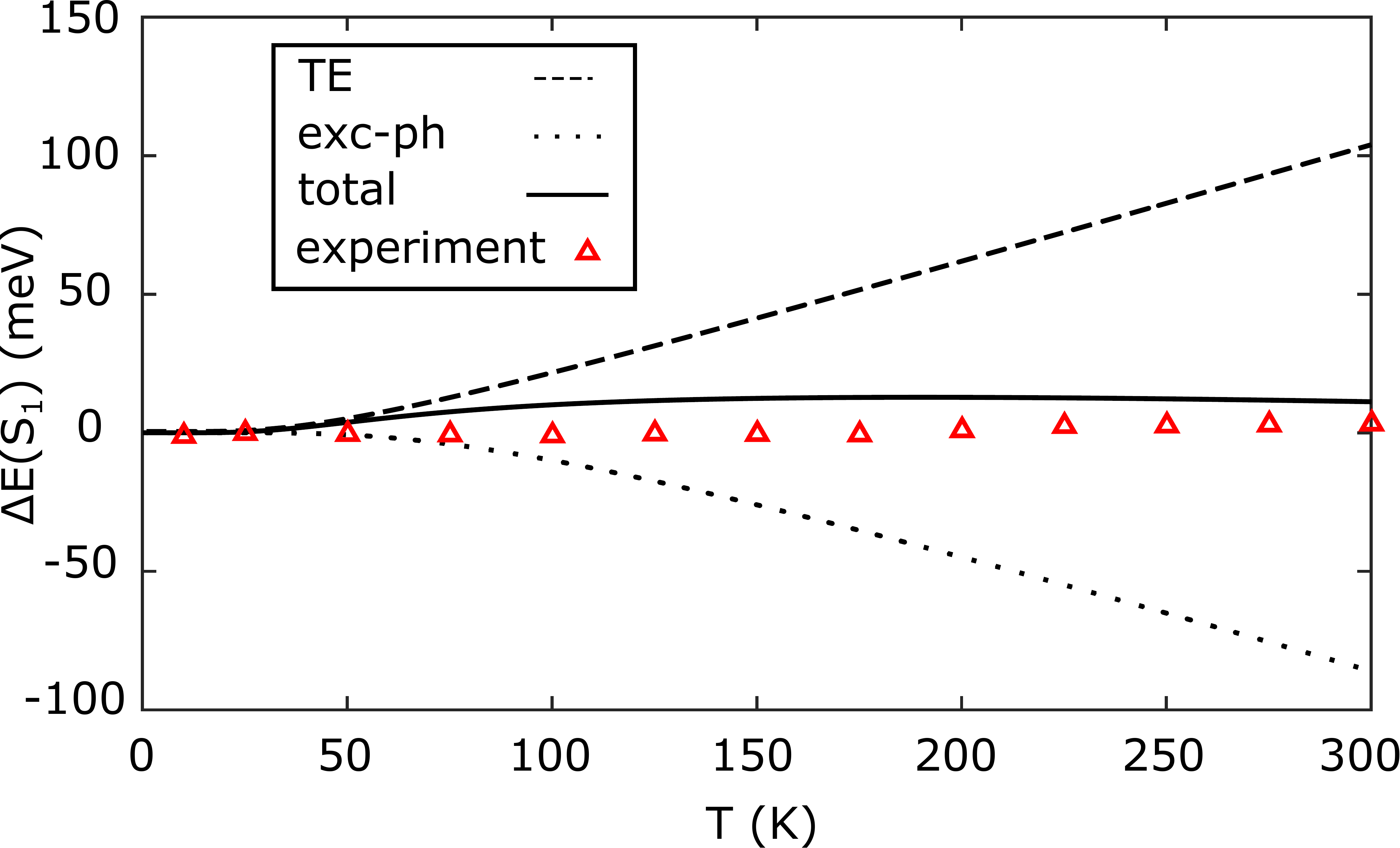}
\caption{Experimental (red-triangles) and theoretical (black solid line) temperature-dependent absorption of pentacene.
%The red triangles indicate the experimental change in the singlet energy relative to its value at $10$\,K, while the black solid line is the theoretical prediction for the same quantity. 
The dashed and dotted lines indicate the individual effects of thermal expansion (TE) and exciton-phonon coupling on the singlet energy.}
%The inset in panel \textbf{b} provides a closer view of the experimental points for anthracene.}
%highlighting the subtle non-monotonic behavior of temperature-dependent absorption.}
\label{fig:temperature}
\end{figure}

As discussed previously and visualized in Figure\,\ref{fig:temperature} for the case of pentacene, the coupling of an exciton to thermally-activated phonons leads to a red shift of its energy (dotted line) relative to its value at $0$\,K. 
%This effect is stronger for pentacene with a more delocalized singlet that depends more sensitively on intermolecular interactions. 
However, exciton-phonon coupling in itself is not enough to capture the experimentally observed temperature dependence of the exciton energy (red triangles in Figure\,\ref{fig:temperature}), and must be complemented by the effect of thermal expansion (TE - dashed line). Thermal expansion increases intermolecular distances, thus approaching the limit of isolated single molecules, and hence an entirely localized Frenkel exciton. This destabilizes the exciton energy, an effect which is more pronounced for delocalized singlets, and less important for states that are relatively localized to begin with. 
%Thus, the blue shift of the pentacene singlet due to thermal expansion is stronger than that of anthracene. 

The effects of thermal expansion and exciton-phonon coupling almost perfectly cancel out, leading to largely temperature-independent exciton energies (solid line), in excellent agreement with experiment. 
%The inset of Figure\,\ref{fig:temperature} provides a closer view of the experimental temperature dependence of anthracene, highlighting the fact that the precise interplay of thermal expansion and exciton-phonon coupling can give rise to complex, non-monotonic behavior. 
Nevertheless, while exciton delocalization determines the magnitude of these two effects individually, the net temperature dependence of the exciton energies is overall weak, something which is also true for tetracene, anthracene and naphthalene, the results for which are given in the SM.
As discussed in detail in section S10 of the SM, the small differences 
between the computational and experimental results,
are mainly due to the challenge associated with accounting for the
effects of very long-wavelength phonons on excitons, as well as the
deviation of some of these modes from the harmonic behavior that our
methodology assumes. We should however emphasize that while individual
phonon modes are treated within the harmonic approximation, our 
methodology still captures exciton-phonon interactions to all orders.
%, as well as phonon-phonon coupling. 

The blue shift of exciton energies due to thermal expansion is caused by the increase in intermolecular distances, while the red shift that results from exciton-phonon coupling is due to the dominance of intraband phonon-induced transitions. Both these effects hold in a wide variety of molecular crystals, unless one considers special cases with anomalous thermal expansion or very small optical gaps, respectively. Therefore, we expect the cancellation of the effects of thermal expansion and exciton-phonon coupling to be present in the vast majority of OSCs, leading to overall weak temperature dependence of exciton energies. 

%\subsection*{Importance of phonon quantum fluctuations for predicting exciton energies}

Since the effect of thermal fluctuations on exciton energies is small, it is quantum fluctuations that are mostly responsible for red shifting their values from those of a static lattice. This leads to a greatly improved agreement between theory and experiment as summarized in Table\,\ref{table:ZPR}, highlighting the predictive power of our approach. This result is particularly important for triplet states, the energy of which is experimentally very challenging to determine \cite{Reineke2014,Ehrler2012}; the correction to the static values for these states is significant due to their highly localized character and associated coupling to the high-frequency modes that are active due to quantum fluctuations. 

The accurate calculation of exciton energies is crucial to various photophysical 
processes in OSCs. If we consider the example of singlet fission, our theoretical framework successfully predicts the experimentally well-established exothermicity ($\text{E}(\text{S}_1)>2 \cdot \text{E}(\text{T}_1)$) of singlet fission in solid pentacene \cite{Rao2011}, within a fully first-principles description. Quantum fluctuations of phonons are crucial to this result, as static approaches using many-body perturbation theory fail to capture it \cite{Refaely-Abramson2017,Liu2020}. At the same time, we also capture the endothermicity of singlet fission in tetracene\cite{Burdett2013}. Generally, the strong renormalization of exciton energies due to quantum fluctuations is an inherent characteristic of OSCs due to the light weight of their constituent elements, and thus the high frequency of oscillations they exhibit. Therefore, accounting for this effect is crucial, regardless of the accuracy of the underlying electronic structure method. 

%The predictive power of our first principles framework will be important for the study of many other applications of OSCs in which precise exciton energies are necessary. As a first example, triplet excitons generated by singlet fission can potentially be transferred to an inorganic material in order to reduce thermalization losses of solar cells, something which was recently achieved for a tetracene-silicon interface \cite{Einzinger2019}. The precise energy difference between the triplets and the bandgap of the inorganic material controls the efficiency of this process. As a second example beyond singlet fission, thermally activated delayed fluorescence in OSCs, a process used to achieve high efficiency in OLED devices \cite{Uoyama2012,Reineke2014}, depends sensitively on the singlet-triplet splitting $\text{E}(\text{S}_1)- \text{E}(\text{T}_1)$.

%The above examples highlight the necessity of achieving highly accurate prediction of exciton energies for a wide variety of optoelectronic applications. 

\begin{table*}[tb]
\centering
  \setlength{\tabcolsep}{4pt} % Default is 6pt
\begin{tabular}{ccccc}
\hline
\hline
$\text{E}(\text{S}_1)$ (eV) &  Calculated static & ZPR (meV) & Calculated with phonons & Experiment \\
\hline
naphthalene & $4.28$ & $-190$ & $4.09$ & $3.90$ \cite{Schnepp1963}  \\
anthracene & $3.44$ & $-160$ & $3.28$ & $3.11$ $- 3.26$ \cite{Schnepp1963} \\
tetracene & $2.45$ & $-90$ & $2.36$ & $2.36$ \cite{Wilson2013} $- 2.39$ \\
pentacene & $1.80$ & $-60$ & $1.74$ & $1.83$ \cite{Rao2011} $- 1.88$  \\
\hline
\hline
$\text{E}(\text{T}_1)$ (eV) &  Calculated static & ZPR (meV) & Calculated with phonons & Experiment \\
\hline
naphthalene & $3.06$ & $-290$ & $2.77$ & $2.63$ \cite{Swiderek1990} \\
anthracene & $2.11$ & $-200$ & $1.91$ & $1.83$ \cite{Reineke2014}\\
tetracene & $1.45$ & $-140$ & $1.31$ & $1.34$ \cite{Reineke2014}\\
pentacene & $0.97$ & $-110$ & $0.86$ & $0.86$ \cite{Rao2010} \\
\hline
\hline
\end{tabular}
\caption{%The effect of zero-point quantum fluctuations of phonons on exciton energies. 
Quantum fluctuations lead to a zero-point renormalization (ZPR) of the exciton energies calculated at the static lattice level, leading to a much improved agreement with experiment. For the singlet energy of anthracene, tetracene and pentacene, we present a range of values, based both on literature values and measurements performed in this work (SM section S9 for experimental spectra).}
\label{table:ZPR}
\end{table*}

%\subsection*{Exciton pressure dependence}

\begin{figure}[tb]
\centering
\includegraphics[width=0.8\linewidth]{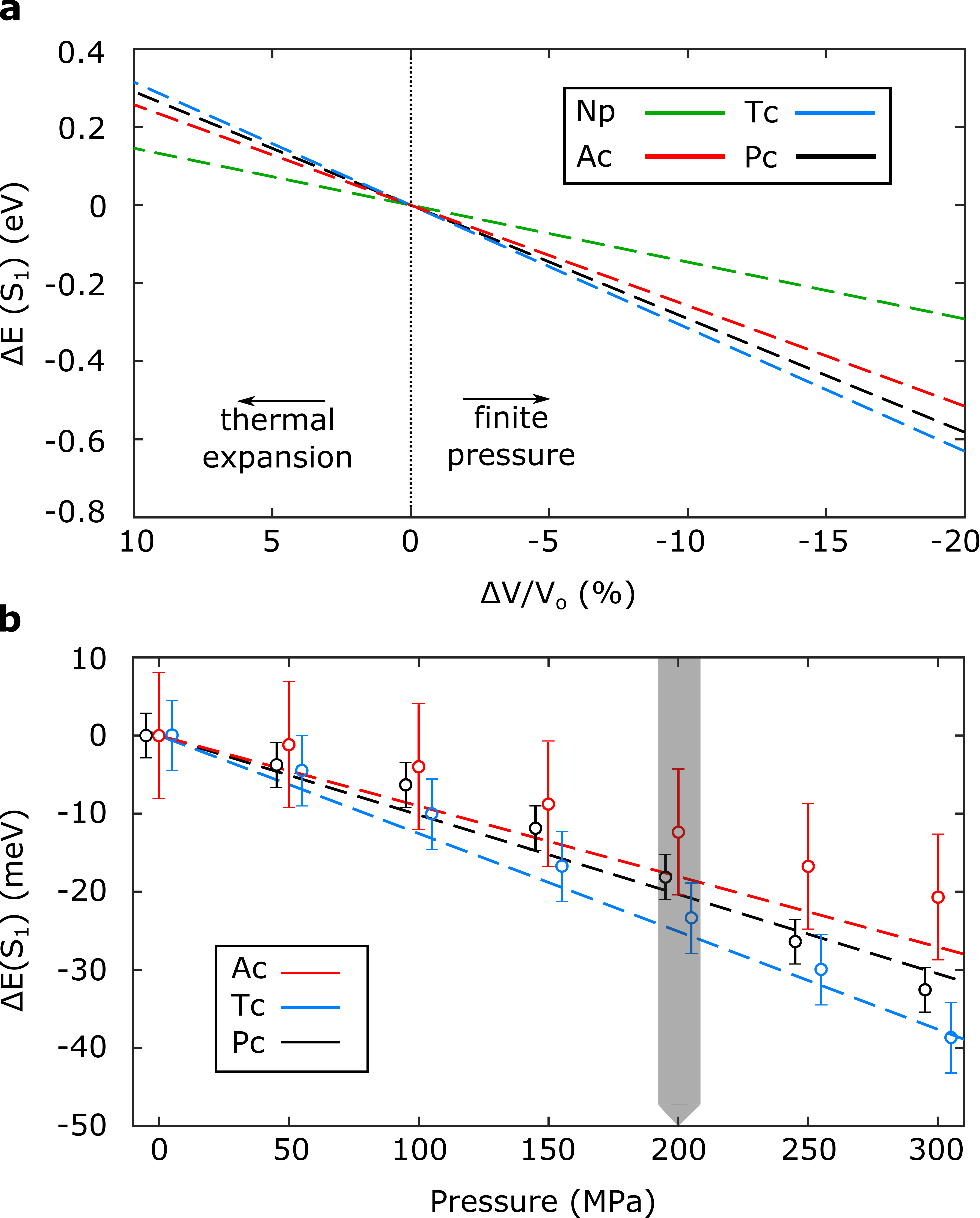}
\caption{Volume dependence of exciton energies. Panel \textbf{a} shows the theoretical prediction over a large range of volume changes. The region with $\Delta V/V_o<0$ corresponds to the application of hydrostatic pressure in the range $0-4$\,GPa.
%More delocalized excitons have a stronger volume dependence, with pentacene being an exception due to subtleties associated with quantum fluctuations as discussed in the text. Thus pressure-dependent absorption can largely be used as a probe of exciton delocalization. 
We compare to pressure-dependent measurements within $0-300$\,MPa in panel \textbf{b}.} %confirming the accuracy of our theoretical framework. 
%The gray shaded area indicates that pressure measurements
%of the different acenes are always carried out at the same pressure values - however, we apply a horizontal shift of $5$\,MPa to the experimental points for visibility purposes.}
\label{fig:pressure}
\end{figure}

The magnitude of the effect of thermal expansion and exciton-phonon coupling on excitons energies depends on the degree of wavefunction delocalization. Unfortunately, the cancellation of these effects that we have demonstrated does not allow one to utilize them as a means of probing delocalization experimentally. Nevertheless, the insights from the above discussion motivate us to examine the effect of pressure on exciton energies. The application of hydrostatic pressure at a given temperature reduces the volume $V_o$ of the unit cell at atmospheric pressure ($\Delta V/V_o<0$), an effect \textit{opposite} to thermal expansion, which increases the unit cell volume ($\Delta V/V_o>0$). However, unlike thermal expansion, the effect of pressure does not compete with phonon-activated processes, leading to a strong red shift of exciton energies \cite{Farina2003}, an effect that we expect to be stronger for states with larger CT character, and could hence be used to probe exciton delocalization. We now examine this effect both from theoretical and experimental points of view. 

Figure~\ref{fig:pressure}a shows the change of the singlet energy of the acene crystals as a function of unit cell volume. The region with $\Delta V/V_o<0$ corresponds to application of finite pressure in the range $0-4$\,GPa. As argued previously, changes in intermolecular distances induced by pressure or thermal expansion have a weaker effect on localized Frenkel-like excitons, but become stronger in states with strong CT character. Indeed, we find that for naphthalene the singlet volume dependence is weaker than for anthracene, which in turn is weaker than that of tetracene. Interestingly, pentacene has a slightly weaker volume dependence than tetracene, which we find is due to qualitative changes in the effect of phonon quantum fluctuations at finite pressures, as discussed in section S3 of the SM. Consistent with this picture, we find in Figure\,S2 that the pressure dependence of the triplet energies is significantly weaker, due to the highly localized nature of these states. In Figure~\ref{fig:pressure}b, we compare our theoretical predictions for the change of the singlet energies at finite pressure to experiment, in the range of $0-300$\,MPa. The full experimental spectra are given in Figure\,S8. We find that theory and experiment are in very good agreement and, remarkably, we correctly predicted the unconventional pressure dependence for pentacene, confirming the accuracy of our theoretical framework.
%, which can also capture the unexpected difference between pentacene and tetracene once we account for the variation of the effect of quantum fluctuations under hydrostatic pressure. 

From the above discussion, it becomes clear that generally, the slope of the experimental exciton energy pressure dependence provides a qualitative measure of exciton delocalization. However, in combination with the experimentally measured exciton temperature dependence, it can also be used to provide an estimate of the magnitude of exciton-phonon interactions due to thermal fluctuations. The shift of exciton energies upon compression $\Delta V/V_o<0$ is linear, and can be extrapolated to $\Delta V/V_o>0$, hence providing an expected blue shift of the exciton energy due to thermal expansion. The difference of this expected blue shift from the experimentally measured energy shift (compared to $0$\,K) gives the magnitude of exciton energy renormalization due to coupling to thermally-activated low-frequency phonons. This is complementary to extracting the magnitude of exciton-phonon interactions from the vibronic progression of the absorption spectrum \cite{Spano2010}, as the latter only provides information on the coupling of the exciton to high-frequency modes.

%\section*{Discussion}

We propose a general framework to study exciton-phonon interactions in organic semiconductors describing localized molecular and extended crystal degrees of freedom simultaneously. We show that exciton delocalization determines the magnitude and nature of these interactions: localized excitons predominantly couple to high-frequency modes via quantum fluctuations, while delocalized excitons interact more strongly with thermally-activated low-frequency phonons. Together with the effect of thermal expansion, which also depends on exciton delocalization, this allows us to reveal the full microscopic mechanism behind the weak temperature of exciton energies in acene crystals, and argue that this should hold in the vast majority of molecular crystals. As a consequence of the weak temperature dependence, the major contribution to exciton energy renormalization compared to the static lattice arises from quantum fluctuations of mostly high-frequency vibrations, always present in organic materials. The magnitude of this renormalization also depends sensitively on exciton delocalization, and accounting for this effect is necessary in order to achieve predictive power for exciton energies.

Overall, our framework provides a unifying picture between the molecular
and crystal limits of organic semiconductors, showing how the delocalization of excitons determines their response to a wide range of structural changes, beyond lattice vibrations and thermal expansion. The effect of pressure provides such an example, and we find that pressure-dependent measurements may be used as an experimental probe of delocalization and thermally-activated exciton-phonon interactions. Therefore, based on factors that determine exciton delocalization, such as spin and the size of molecules, one can anticipate the difference in the response of different materials to a variety of structural changes.

\noindent
%Tetracene and pentacene powders were bought from Sigma Aldrich (99\% 
%purity) and used as supplied. The powder is transferred to an evaporator
%chamber (Kurt J Lesker) and kept overnight under high vacuum ($< 
%10^{-6}$\,mbar) to remove any residual solvent prior to thermal 
%evaporation. A $150$-nm thick layer of pentacene/tetracene is then 
%thermally evaporated at a rate of $0.5\pm 0.1$ \si{\angstrom}/s. The rate
%is monitored using a standard Quartz-crystal microbalance and calibrated
%with atomic force microscopy measurements.
%Anthrancene (reagent grade 97\%) was purchased from Sigma Aldrich and 
%used as supplied. Crystallites of anthracene are dissolved in 
%choloroform (HPLC grade, Sigma Aldrich) to make a saturated solution at 
%$50$\,\textsuperscript{o}C. The hot solution is then spin coat at 
%$1000$\,rpm to form a homogenous film on the fused silica substrate. 
%The low sublimation point of naphthalene ($80$\,\textsuperscript{o}C) 
%prevents both thermal evaporation or solution processing of this 
%material such that reliable optical measurements can be performed. 

%\textbf{Temperature-dependent absorption}

\noindent
%An Agilent Cary 6000i UV–vis–NIR spectrophotometer with blank substrate 
%correction is used for all measurements. Samples prepared on fused 
%silica substrates are placed in a continuous-flow cryostat (Oxford 
%Instruments Optistat CF-V) under an argon atmosphere. We allow the 
%sample temperature to equilibrate for $30$\,min before taking data. Measurements were taken with $1$\,nm wavelength steps between $250 – 900$\,nm, with $1$\,s of integration at each wavelength.

%\textbf{Pressure-dependent measurements}

%Transmittance spectra of the organic molecules are measured with a LAMBDA 750 UV/Vis/NIR Spectrophotometer (Perkin Elmer). The samples are kept inside a high-pressure cell (ISS Inc.) filled with an inert liquid, Fluorinert FC-72 (3M). Hydrostatic pressure is generated through a pressurizing liquid using a manual pump. Prior using, the liquid is degassed in a Schlenk line to remove oxygen which causes, from $300$\,MPa onwards, scattering of a fraction of light and therefore a reduction of the transmitted signal from the sample. The pressure is applied from ambient pressure to $300$\,MPa in steps of $50$\,MPa. Before the measurement, we wait $7$ minutes for equilibration of the material under pressure. We estimate an error of the pressure reading to be $20$\,MPa.

%\section*{Data availability}
%The data underlying all figures in the main text and supplementary information are publicly available at [URL added in proof].

%\section*{Code availability}
%The custom code used for phonon and exciton-phonon calculations is available by written request to bm418@cam.ac.uk.

A.M.A. is thankful to Jeffrey B. Neaton, Jonah B. Haber (UC Berkeley),
Felipe H. da Jornada (Stanford) and Cristoph Schnedermann (Cambridge) for
insightful discussions. The authors 
acknowledge the support of the Winton Programme for the Physics of 
Sustainability. A.M.A. acknowledges the support of the Engineering and 
Physical Sciences Research Council (EPSRC) for funding under grant 
EP/L015552/1. The work of L.A.M. and B.E. is part of the Dutch Research Council (NWO) and was performed at the research institute AMOLF, supported by NWO Vidi grant 016.Vidi.179.005. R.H.F. acknowledges the Simons Foundation grant 601946. B.M. 
acknowledges support from the Gianna Angelopoulos Programme for Science, Technology, and Innovation. Part of the 
calculations were performed using resources provided by the Cambridge Tier-2
system operated by the University of Cambridge Research Computing Service 
(http://www.hpc.cam.ac.uk) and funded by EPSRC Tier-2 capital grant 
EP/P020259/1.

%\section*{Author contributions}
%A.M.A. and B.M. conceived of the study. All calculations were
%performed by A.M.A., under guidance from B.M. R.P. and M.N. performed the
%temperature-dependent measurements, samples were prepared by N.S. L.A.M.
%and B.E. performed the pressure-dependent measurements. A.M.A., B.M., A.R., R.H.F. and %A.W.C. interpreted the computational and experimental 
%results. A.M.A. and B.M. wrote
%the manuscript with input from all the authors.

\textbf{}
\bibliographystyle{unsrt}
\bibliography{references}

\begin{thebibliography}{10}

\bibitem{Reineke2009}
Sebastian Reineke, Frank Lindner, Gregor Schwartz, Nico Seidler, Karsten
  Walzer, Bj{\"{o}}rn L{\"{u}}ssem, and Karl Leo.
\newblock {White organic light-emitting diodes with fluorescent tube
  efficiency.}
\newblock {\em Nature}, 459(7244):234--8, 2009.

\bibitem{Meng2018}
Lingxian Meng, Yamin Zhang, Xiangjian Wan, Chenxi Li, Xin Zhang, Yanbo Wang,
  Xin Ke, Zuo Xiao, Liming Ding, Ruoxi Xia, Hin-Lap Yip, Yong Cao, and
  Yongsheng Chen.
\newblock Organic and solution-processed tandem solar cells with 17.3\%
  efficiency.
\newblock {\em Science}, 361(6407):1094--1098, 2018.

\bibitem{Knupfer2003}
M.~Knupfer.
\newblock Exciton binding energies in organic semiconductors.
\newblock {\em Applied Physics A}, 77(5):623--626, Oct 2003.

\bibitem{Kohler2009}
A.~K{\"{o}}hler and H.~B{\"{a}}ssler.
\newblock {Triplet states in organic semiconductors}.
\newblock {\em Materials Science and Engineering R: Reports}, 66(4-6):71--109,
  2009.

\bibitem{Uoyama2012}
Hiroki Uoyama, Kenichi Goushi, Katsuyuki Shizu, Hiroko Nomura, and Chihaya
  Adachi.
\newblock {Highly efficient organic light-emitting diodes from delayed
  fluorescence}.
\newblock {\em Nature}, 492:234, December 2012.

\bibitem{Reineke2014}
Sebastian Reineke.
\newblock {Phosphorescence meets its match}.
\newblock {\em Nature Photonics}, 8:269, March 2014.

\bibitem{Smith2010}
Millicent~B. Smith and Josef Michl.
\newblock {Singlet fission}.
\newblock {\em Chemical Reviews}, 110(11):6891--6936, 2010.

\bibitem{Hanna2006}
M.~C. Hanna and A.~J. Nozik.
\newblock {Solar conversion efficiency of photovoltaic and photoelectrolysis
  cells with carrier multiplication absorbers}.
\newblock {\em Journal of Applied Physics}, 100(7), 2006.

\bibitem{Rao2017}
Akshay Rao and Richard~H Friend.
\newblock {Harnessing singlet exciton fission to break the Shockley–Queisser
  limit}.
\newblock {\em Nature Reviews Materials}, 2:17063, October 2017.

\bibitem{Beljonne2013}
D.~Beljonne, H.~Yamagata, J.~L. Br{\'{e}}das, F.~C. Spano, and Y.~Olivier.
\newblock {Charge-transfer excitations steer the davydov splitting and mediate
  singlet exciton fission in pentacene}.
\newblock {\em Physical Review Letters}, 110(22):1--5, 2013.

\bibitem{Berkelbach2013}
Timothy~C. Berkelbach, Mark~S. Hybertsen, and David~R. Reichman.
\newblock {Microscopic theory of singlet exciton fission. II. Application to
  pentacene dimers and the role of superexchange}.
\newblock {\em Journal of Chemical Physics}, 138(11), 2013.

\bibitem{Yost2014}
Shane~R Yost, Jiye Lee, Mark W~B Wilson, Tony Wu, David~P McMahon, Rebecca~R
  Parkhurst, Nicholas~J Thompson, Daniel~N Congreve, Akshay Rao, Kerr Johnson,
  Matthew~Y Sfeir, Moungi~G Bawendi, Timothy~M Swager, Richard~H Friend, Marc~a
  Baldo, and Troy {Van Voorhis}.
\newblock {A transferable model for singlet-fission kinetics}.
\newblock {\em Nature Chemistry}, 6(6):492--497, 2014.

\bibitem{Coto2015}
Pedro~B. Coto, Sahar Sharifzadeh, Jeffrey~B. Neaton, and Michael Thoss.
\newblock {Low-lying electronic excited states of pentacene oligomers: A
  comparative electronic structure study in the context of singlet fission}.
\newblock {\em Journal of Chemical Theory and Computation}, 11(1):147--156,
  2015.

\bibitem{Cudazzo2015}
Pierluigi Cudazzo, Francesco Sottile, Angel Rubio, and Matteo Gatti.
\newblock {Exciton dispersion in molecular solids}.
\newblock {\em Journal of Physics Condensed Matter}, 27(11), 2015.

\bibitem{Refaely-Abramson2017}
Sivan Refaely-Abramson, Felipe~H. {Da Jornada}, Steven~G. Louie, and Jeffrey~B.
  Neaton.
\newblock {Origins of Singlet Fission in Solid Pentacene from an ab initio
  Green's Function Approach}.
\newblock {\em Physical Review Letters}, 119(26):1--6, 2017.

\bibitem{Arago2015}
Juan Arag{\'{o}} and Alessandro Troisi.
\newblock {Dynamics of the excitonic coupling in organic crystals}.
\newblock {\em Physical Review Letters}, 114(2):1--5, 2015.

\bibitem{Musser2015}
Andrew~J Musser, Matz Liebel, Christoph Schnedermann, Torsten Wende, Tom~B
  Kehoe, Akshay Rao, and Philipp Kukura.
\newblock {Evidence for conical intersection dynamics mediating ultrafast
  singlet exciton fission}.
\newblock {\em Nature Physics}, 11:352, March 2015.

\bibitem{Bakulin2016}
Artem~A. Bakulin, Sarah~E. Morgan, Tom~B. Kehoe, Mark~W.B. Wilson, Alex~W.
  Chin, Donatas Zigmantas, Dassia Egorova, and Akshay Rao.
\newblock {Real-time observation of multiexcitonic states in ultrafast singlet
  fission using coherent 2D electronic spectroscopy}.
\newblock {\em Nature Chemistry}, 8(1):16--23, 2016.

\bibitem{Miyata2017}
Kiyoshi Miyata, Yuki Kurashige, Kazuya Watanabe, Toshiki Sugimoto, Shota
  Takahashi, Shunsuke Tanaka, Jun Takeya, Takeshi Yanai, and Yoshiyasu
  Matsumoto.
\newblock {Coherent singlet fission activated by symmetry breaking}.
\newblock {\em Nature Chemistry}, 9(10):983--989, 2017.

\bibitem{Schnedermann2019}
Christoph Schnedermann, Antonios~M Alvertis, Torsten Wende, Steven Lukman,
  Jiaqi Feng, Florian A Y~N Schr{\"{o}}der, David H~P Turban, Jishan Wu,
  Nicholas D~M Hine, Neil~C Greenham, Alex~W Chin, Akshay Rao, Philipp Kukura,
  and Andrew~J Musser.
\newblock {A molecular movie of ultrafast singlet fission}.
\newblock {\em Nature Communications}, 10:4207, 2019.

\bibitem{Alvertis2019}
Antonios~M. Alvertis, Florian~A.Y.N. Schr{\"{o}}der, and Alex~W. Chin.
\newblock {Non-equilibrium relaxation of hot states in organic semiconductors:
  Impact of mode-selective excitation on charge transfer}.
\newblock {\em Journal of Chemical Physics}, 151(8), 2019.

\bibitem{Troisi2005}
Alessandro Troisi, Giorgio Orlandi, and John~E. Anthony.
\newblock {Electronic interactions and thermal disorder in molecular crystals
  containing cofacial pentacene units}.
\newblock {\em Chemistry of Materials}, 17(20):5024--5031, 2005.

\bibitem{Troisi2006}
Alessandro Troisi and Giorgio Orlandi.
\newblock {Dynamics of the intermolecular transfer integral in crystalline
  organic semiconductors}.
\newblock {\em Journal of Physical Chemistry A}, 110(11):4065--4070, 2006.

\bibitem{Lars1965}
Hedin Lars.
\newblock {New Method for Calculating the One-Particle Green's Function with
  Application to the Electron-Gas Problem}.
\newblock {\em Physical Review}, 139(17):796--823, 1965.

\bibitem{Rohlfing1998}
Michael Rohlfing and Steven~G. Louie.
\newblock {Electron-hole excitations in semiconductors and insulators}.
\newblock {\em Physical Review Letters}, 81(11):2312--2315, 1998.

\bibitem{Rohlfing2000}
Michael Rohlfing and Steven~G. Louie.
\newblock {Electron-hole excitations and optical spectra from first
  principles}.
\newblock {\em Physical Review B - Condensed Matter and Materials Physics},
  62(8):4927--4944, 2000.

\bibitem{Kunc1982}
K.~Kunc and Richard~M. Martin.
\newblock {Ab initio force constants of {G}a{A}s: A new approach to calculation
  of phonons and dielectric properties}.
\newblock {\em Physical Review Letters}, 48(6):406--409, 1982.

\bibitem{Monserrat2018}
Bartomeu Monserrat.
\newblock {Electron – phonon coupling from finite differences}.
\newblock {\em Journal of Physics Condensed Matter}, 30, 2018.

\bibitem{Monserrat2016_2}
Bartomeu Monserrat.
\newblock {Vibrational averages along thermal lines}.
\newblock {\em Physical Review B}, 93(1):1--10, 2016.

\bibitem{Ehrler2012}
Bruno Ehrler, Brian~J. Walker, Marcus~L. B{\"{o}}hm, Mark~W.B. Wilson, Yana
  Vaynzof, Richard~H. Friend, and Neil~C. Greenham.
\newblock {In situ measurement of exciton energy in hybrid singlet-fission
  solar cells}.
\newblock {\em Nature Communications}, 3(May), 2012.

\bibitem{Saha2014}
Kush Saha and Ion Garate.
\newblock {Phonon-induced topological insulation}.
\newblock {\em Physical Review B - Condensed Matter and Materials Physics},
  89(20):1--13, 2014.

\bibitem{Rao2011}
Akshay Rao, Mark~W.B. Wilson, Sebastian Albert-Seifried, Riccardo {Di Pietro},
  and Richard~H. Friend.
\newblock {Photophysics of pentacene thin films: The role of exciton fission
  and heating effects}.
\newblock {\em Physical Review B - Condensed Matter and Materials Physics},
  84(19):1--8, 2011.

\bibitem{Liu2020}
Xingyu Liu, Rithwik Tom, Xiaopeng Wang, Cameron Cook, Bohdan Schnatschneider,
  and Noa Marom.
\newblock {Pyrene-Stabilized Acenes as Intermolecular Singlet Fission
  Candidates: Importance of Exciton Wave-Function Convergence}.
\newblock {\em Journal of Physics Condensed Matter}, 2020.

\bibitem{Burdett2013}
Jonathan~J. Burdett and Christopher~J. Bardeen.
\newblock {The dynamics of singlet fission in crystalline tetracene and
  covalent analogs}.
\newblock {\em Accounts of Chemical Research}, 46(6):1312--1320, 2013.

\bibitem{Schnepp1963}
O~Schnepp.
\newblock {Electronic Spectra of Molecular Crystals}.
\newblock {\em Annual Review of Physical Chemistry}, 14(1):35--60, 1963.

\bibitem{Wilson2013}
Mark~W.B. Wilson, Akshay Rao, Kerr Johnson, Simon G{\'{e}}linas, Riccardo {Di
  Pietro}, Jenny Clark, and Richard~H. Friend.
\newblock {Temperature-independent singlet exciton fission in tetracene}.
\newblock {\em Journal of the American Chemical Society}, 135(44):16680--16688,
  2013.

\bibitem{Swiderek1990}
Petra Swiderek, M.~Michaud, G.~Hohlneicher, and L.~Sanche.
\newblock {Electron energy loss spectroscopy of solid naphthalene and
  acenaphthene: search for the low-lying triplet states}.
\newblock {\em Chemical Physics Letters}, 175(6):667--673, 1990.

\bibitem{Rao2010}
Akshay Rao, Mark~W.B. Wilson, Justin~M. Hodgkiss, Sebastian Albert-Seifried,
  Heinz B{\"{a}}ssler, and Richard~H. Friend.
\newblock {Exciton Fission and Charge Generation via Triplet Excitons in
  Pentacene/$\text{C}_{60}$ Bilayers}.
\newblock {\em Journal of the American Chemical Society}, 132(36):12698--12703,
  2010.

\bibitem{Farina2003}
L.~Farina, K.~Syassen, Aldo Brillante, R.~G. {Della Valle}, E.~Venuti, and
  N.~Karl.
\newblock {Pentacene at high pressure}.
\newblock {\em High Pressure Research}, 23(3 SPEC. ISS.):349--354, 2003.

\bibitem{Spano2010}
Frank~C. Spano.
\newblock {The spectral signatures of frenkel polarons in H- And J-aggregates}.
\newblock {\em Accounts of Chemical Research}, 43(3):429--439, 2010.

\end{thebibliography}

\end{document}